# Photocurrent Imaging of Multi-Memristive Charge Density Wave Switching in Two-Dimensional 1T-TaS$_2$

Tarun Patel[1], Junichi Okamoto[2], Tina Dekker[1], Bowen Yang[1], Jingjing Gao[3,4], Xuan Luo[3], Wenjian Lu[3], Yuping Sun[3,5,6], and Adam W. Tsen[1*]

[1]Institute for Quantum Computing, Department of Physics and Astronomy, Department of Electrical and Computer Engineering, and Department of Chemistry, University of Waterloo, Waterloo, Ontario N2L 3G1, Canada

[2]Institute of Physics, University of Freiburg, D-79104 Freiburg, Germany

[3]Key Laboratory of Materials Physics, Institute of Solid State Physics, Chinese Academy of Sciences, Hefei 230031, People's Republic of China

[4]University of Science and Technology of China, Hefei 230026, People's Republic of China

[5]High Magnetic Field Laboratory, Chinese Academy of Sciences, Hefei 230031, People's Republic of China

[6]Collaborative Innovation Centre of Advanced Microstructures, Nanjing University, Nanjing 210093, People's Republic of China

*Correspondence to: awtsen@uwaterloo.ca

**Abstract:**

Transport studies of atomically thin 1T-TaS$_2$ have demonstrated the presence of intermediate resistance states across the nearly commensurate (NC) to commensurate (C) charge density wave (CDW) transition, which can be further switched electrically. While this presents exciting opportunities for the material in memristor applications, the switching mechanism has remained elusive and could be potentially attributed to the formation of inhomogeneous C and NC domains across the 1T-TaS$_2$ flake. Here, we present simultaneous electrical driving and scanning photocurrent imaging of CDWs in ultrathin 1T-TaS$_2$ using a vertical heterostructure geometry. While micron-sized CDW domains form upon changing temperature, electrically driven transitions result in largely uniform changes, indicating that states of intermediate resistance for the latter likely correspond to true metastable CDW states in between the NC and C phases, which we then explain by a free energy analysis. Additionally, we are able to perform repeatable and bidirectional switching across the multiple CDW states without changing sample temperature, demonstrating that atomically thin 1T-TaS$_2$ can be further used as a robust and reversible multi-memristor material.

**Main Text:**

Memristors exhibit different states of electrical resistance that can be hysteretically switched by voltage or current. Such devices hold promise for potential applications in neuromorphic computing and data storage[1]. Phase-change materials, which display different intrinsic resistivities across a first-order phase transition, have attracted particular attention as candidates for memristors due to their fast switching speeds and robust behavior[2]. Recently, several groups have demonstrated memristive switching in devices incorporating the layered, charge density wave (CDW) material, 1T-TaS$_2$[3–9]. Upon warming from the insulating, commensurate (C) CDW phase, the resistivity of bulk(-like) 1T-TaS$_2$ abruptly decreases by an order of magnitude at ~220K upon entering the nearly commensurate (NC) CDW phase, which consists of ~10nm-sized C-phase regions separated by a network of more conductive discommensuration channels[10–14]. At constant temperature, both volatile and nonvolatile transitions between these two states have been induced electrically by DC and pulsed voltages/currents[3,6]. In ultrathin flakes, metastable states with intermediate resistance levels have further been observed under equilibrium conditions[4,5], although the switching behavior demonstrated so far has only been unidirectional at a given temperature.

The multiple CDW states are formed intrinsically in two-dimensional (2D) 1T-TaS$_2$, which possesses a greater activation barrier between the C and NC phases due to enhanced pinning of discommensurations[5,15]. They can be potentially exploited for enhanced memristor functionalities[16]; however, two major challenges need to be addressed. First, the nature of the metastable states remains elusive. One possibility is that they simply arise from inhomogeneous C and NC domains formed across the flake (on a length scale much larger than the natural NC discommensuration spacing), which would make device miniaturization less appealing. Second, it is important to be able to switch between the higher and lower resistance levels back and forth bidirectionally without externally changing sample temperature, which is a slow and cumbersome process. In this work, we resolve both these issues through a scanning photocurrent study of ultrathin 1T-TaS$_2$ devices incorporated in a novel heterostructure geometry. This scheme allows us to differentiate between the C and NC phases (as well as the metastable states in between) spatially on the micron scale across different temperatures and electrical driving conditions. While changing temperature results in the clear formation of large CDW domains, electrically induced transitions appear more uniform and can be modeled using classical discommensuration theory. In addition, we uncover a parameter space where the multistep transitions can be bidirectionally switched repeatedly at a fixed temperature, allowing for true multi-memristive functionality in a 2D material for the first time.

A schematic of our unique device and measurement scheme is shown in Fig. 1a. A scanning, focused laser (532nm wavelength) impinges on a heterostructure junction consisting of (from top to bottom) 1T-TaS$_2$, WSe$_2$, and graphite (Gr) thin flakes. The detailed geometry and fabrication process can be found in the Supporting Information, Materials and Methods. In the absence of light, the resistance across the junction is relatively large (≳100MΩ). With light, electron-hole pairs generated in semiconducting WSe$_2$ will separate due to the intrinsic electric field created by the work function mismatch between 1T-TaS$_2$ (5.2eV) and graphite (4.6eV)[17,18]. This process then yields a local, vertical photocurrent, $I_{pc}$, even when no bias is applied to the junction (see Supporting Information, Fig. S1, for full current-voltage characteristics with and without light).

Across the NC-C transition, the Fermi surface of 1T-TaS$_2$ undergoes a reconstruction[19], which will manifest as a change in $I_{pc}$. Figure 1b shows a band structure diagram for this process. Scanning the laser across our sample allows for the C and NC domains (as well as intermediate metastable states) to be imaged spatially with diffraction-limited resolution (~1μm). The 1T-TaS$_2$ is further contacted with two leads laterally outside the junction area, so that its resistance (order ~10kΩ) can be monitored and the NC-C transition be driven electrically like that in previous works[4,5]. We note that this geometry stands in contrast to more common planar, two-terminal photocurrent devices[20–22], which we have found to yield less direct information on CDW domain structure in 1T-TaS$_2$ (see Supporting Information, Fig S2).

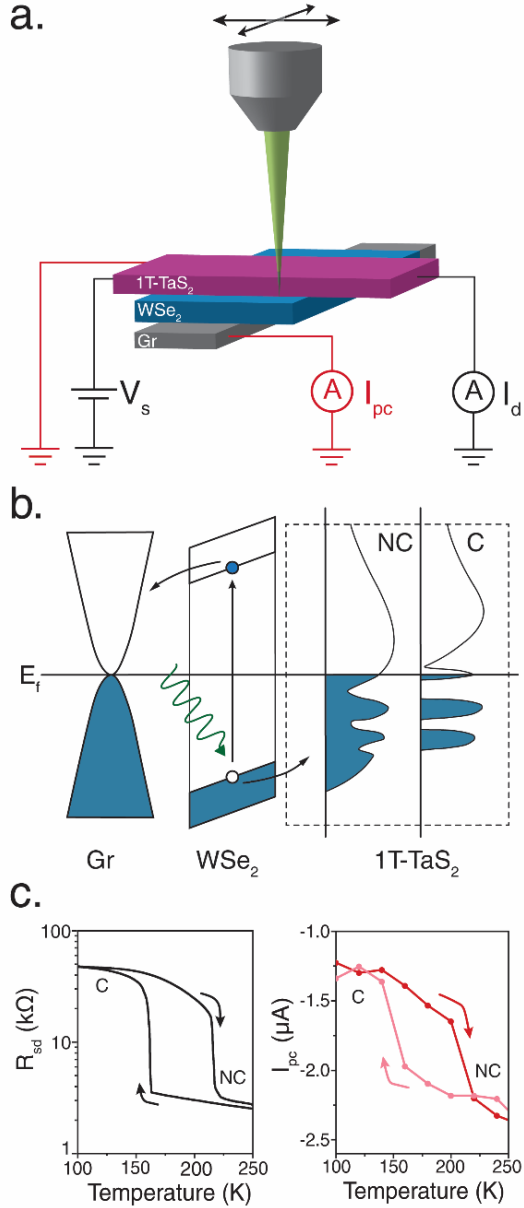

**Figure 1.** Simultaneous electrical driving and photocurrent detection of NC-C CDW transition in 2D 1T-TaS$_2$. (a) Schematic of device heterostructure and measurement geometry. Lateral contacts to 1T-TaS$_2$ allow measurement of resistance and electrical driving, while vertical contacts give photocurrent at zero bias. (b) Band structure diagram showing mechanism of photocurrent generation. Photo-generated electron-hole pairs in WSe$_2$ separate due to the intrinsic electric field created by the work function mismatch between 1T-TaS$_2$ and graphite, which changes across the NC-C transition. (c) Comparison between temperature-dependent resistance of a 22nm-thick 1T-TaS$_2$ device (left) and photocurrent signal with a defocused laser. Both track the hysteretic NC-C transition.

In order to demonstrate the working principle described above, we first compare in Fig. 1c the temperature-dependent lateral resistance, $R_{sd} = V_s/I_d$, for a device with a moderate 1T-TaS$_2$ thickness of 22nm, to that of photocurrent $I_{pc}$, with laser defocused across the entire junction. Both show the abrupt and hysteretic NC-C transition centered at ~190K that is characteristic of bulk-like 1T-TaS$_2$, indicating the validity of our measurement scheme. The negative sign of $I_{pc}$ indicates that photogenerated electrons (holes) flow into Gr (1T-TaS$_2$), consistent with the work function difference already mentioned.

We now turn to photocurrent imaging of the temperature-driven transition. Figure 2a shows an optical image of an ultrathin device with 5nm 1T-TaS$_2$ thickness. The flake has a discontinuity at the narrowest region (see Supporting Information, Fig. S3). Photocurrent images of the junction area (with both contacts to 1T-TaS$_2$ grounded) are shown in Fig. 2b as a function of temperature for both cooling and warming around the transition temperature. Red (blue) regions where $I_{pc}$ is larger (smaller) in magnitude correspond to those more in the NC (C) phase. At higher temperatures, the entire area is relatively uniform and in the NC phase, although there is a marked difference across the discontinuity. The spot-like features are attributed to small bubbles or unwanted particles at the interface and likely do not reflect true features of the CDWs (see Supporting Information, Fig. S4).

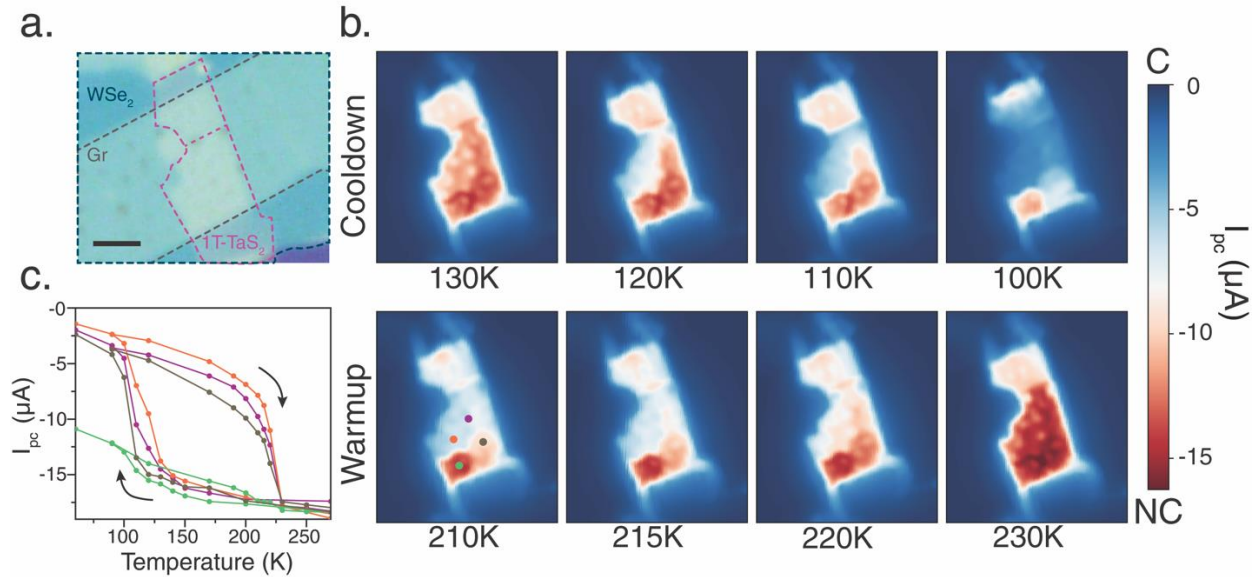

**Figure 2.** Photocurrent imaging of the temperature-driven NC-C transition. (a) Optical image of a device with 5nm-thick 1T-TaS$_2$ with flake edges traced out for clarity. Scale bar is 5μm. (b) Photocurrent maps taken at various temperatures across the NC-C transition for cooling and warming show the formation of micron-sized domains. (c) Temperature-dependent photocurrent signal from four different color-coded locations marked in the bottom-left image in (b). Each region shows distinct transition temperatures and metastability.

At lower temperatures, different areas of the junction transition into the C phase at different temperatures and to varying degrees, manifesting in the formation of micron-sized domains in the photocurrent image. In order to show the regional differences more explicitly, in Fig. 2c we have plotted the temperature-dependent photocurrent values at four different locations marked by the colored points in the bottom left image in Fig. 2b. Overall, the hysteresis between cooling and warming is wider than that for the thicker sample shown in Fig. 1c, which is consistent with previous transport studies[5]. Upon cooling, the temperature at which the transition into the C phase begins varies over 25K across the four locations, while the abruptness of the transition is also different. In particular, orange and purple traces show several metastable states with intermediate $I_{pc}$ that is absent in the gray trace. In contrast, upon warming, all locations complete the transition into the NC phase at ~225K. Interestingly, the location marked in green maintains a relatively large $I_{pc}$ throughout the entire temperature range, indicating that it remains nearly locked in the NC phase, although a small hysteresis loop is still visible. Such an effect has also been seen in transport measurements on ultrathin samples, although no spatial information was obtained[5].

For comparison, we have repeated this measurement on the 22-nm-thick $1T-TaS_2$ device, which shows a narrower hysteresis, more abrupt NC-C transition, and more complete transition into the C phase across the entire junction (see Supporting Information, Fig. S5). This is also consistent with previous nanoscale infrared imaging on a $1T-TaS_2$ flake with moderate thickness[23]. Taken together, these measurements demonstrate the utility of our photocurrent scheme in deciphering spatial differences in the NC-C transition of $1T-TaS_2$ on the micron scale that is relevant for electronic devices.

We thus turn to the electrically induced NC-C transitions. In the upper panel of Fig. 3a, we show a series of slow, lateral current-voltage sweeps (0.13V/s maximum) taken on another $1T-TaS_2$ device with thickness 5-8nm starting in the NC phase at 170K. Initially, $I_d$ increases linearly with applied $V_s$ until a sudden decrease is observed at a relatively small voltage value of ~1.4V, after which we sweep $V_s$ back to zero. As the measurement is repeated, the conductance decreases after each subsequent sweep while the current-voltage characteristics become increasingly nonlinear. This process gradually transitions the $1T-TaS_2$ sample further into the C phase with higher resistivity and nonlinear conduction[24]. The voltage needed to drive the current decrease increases with the sweep number, although the $I_d$-$V_s$ characteristics appear to eventually saturate onto a single curve (see sweep 9). In the upper panel of Fig. 3c, we have explicitly plotted $R_{sd}$ at low bias with sweep number before each sweep, which also begins to show a saturation after sweep 9. We note that this saturated resistance level is 63% of the resistance of the C phase driven by temperature at 170K, indicating that conducting discommensurations still remain.

In this saturated, metastable state close to the C phase, increasing $V_s$ further to a larger value of ~11V at the same temperature of 170K induces an abrupt current rise, as can be seen in sweep 12 in the lower panel of Fig. 3a. Upon sweeping $V_s$ back to zero, the current abruptly decreases at a lower voltage of ~8.5V, but does not return to its original state. As a result, the device can be switched back towards the NC state with higher conductance in a nonvolatile manner. We have repeated this driving procedure with additional sweeps and the corresponding resistances are shown in the upper panel of Fig. 3c, which demonstrates that 2D $1T-TaS_2$ can be switched bidirectionally back and forth between the NC- and C-like states at a fixed temperature.

Such electrically induced effects could potentially arise from the formation and switching of micron-sized NC or C domains, as in the case of the temperature-driven transition shown in Fig. 2. To address this issue, we further performed zero-bias photocurrent scans before many of the voltage sweeps, and five selected images are shown in Fig. 3b. The overall magnitude of $I_{pc}$ follows the device conductance—decreasing from sweep 1 to 6 to 9, increasing from 9 to 12, and decreasing again from 12 to 13. While the images are not completely homogeneous, they do not show the characteristic domains formed during cooling or warming. To show this more explicitly, we have tracked $I_{pc}$ at five different locations across the junction (see colored points in the lower image of Fig. 3c), and their normalized change with drive number is shown in the lower panel of Fig. 3c, with the different traces offset for clarity. With the exception of the orange trace, which shows slightly more saturated characteristics after sweep 9, the photocurrent change for all other locations uniformly follows the overall 1T-TaS$_2$ resistance. This indicates that if CDW domains were created during the electrical driving, their length scale should be much smaller than 1μm, the laser spot size.

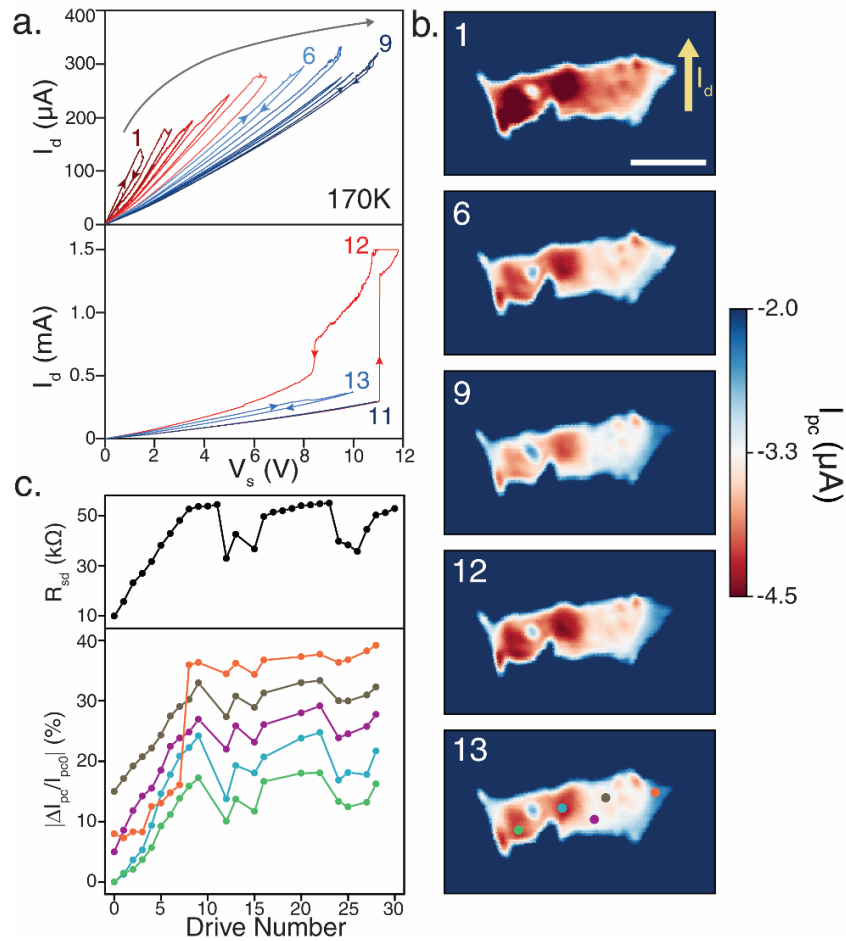

**Figure 3.** Reversible electrical driving of metastable states across NC-C transition and concurrent photocurrent imaging. (a) Sequential current-voltage sweeps taken at 170K starting in the NC phase on a 5-8nm-thick 1T-TaS$_2$ device. The top panel shows a gradual decrease in conductance from sweeps 1-9. The bottom panel shows an abrupt increase in conductance in sweep 12 and decrease again for sweep 13. (b) Photocurrent maps taken after selected sweeps show changes following the resistance trend without domain formation. Scale bar is 5μm. (c) Top: lateral 1T-TaS$_2$ resistance before each voltage sweep measured at low bias. The multiple resistance states can be switched back and forth repeatedly at a fixed temperature. Bottom: photocurrent change (in percent) for various color-coded locations marked in the bottom image in (b). The traces have been offset for clarity. All locations show photocurrent tracking the overall resistance.

We now consider the potential mechanisms for the observed homogeneous switching. While the NC-to-C transition proceeds in multiple steps, the reverse transition is abrupt. This likely reflects different origins for the two, and so we discuss the forward and backward transitions separately. The former has been previously seen in ultrathin flakes, but has not been explained[4,5]. To this end, we have constructed a free energy model based on classical discommensuration theory to account for this transition that is consistent with our new spatially resolved findings. We describe these results in brief below, deferring the full analysis to the Supporting Information.

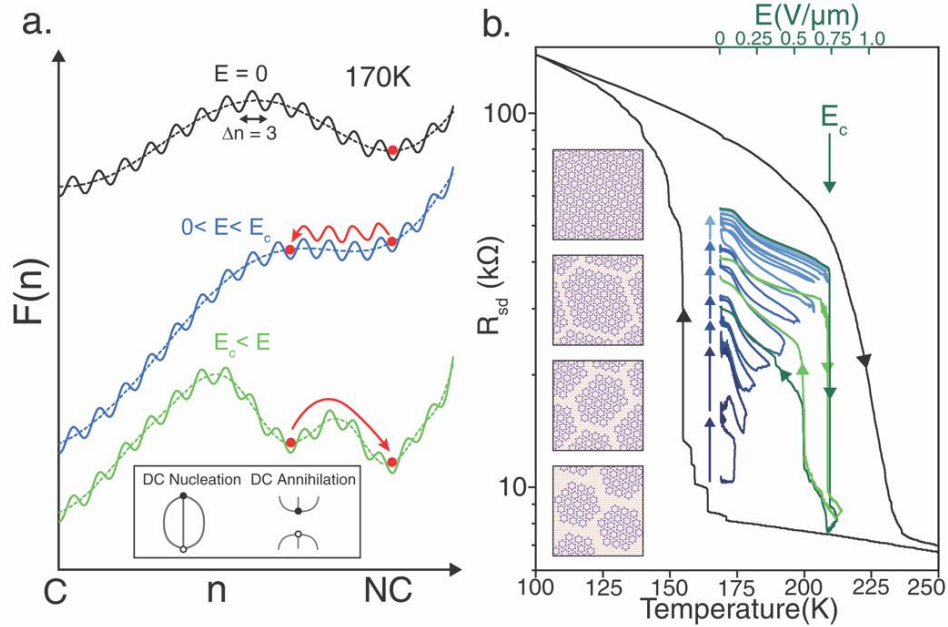

**Figure 4.** Proposed mechanism for bidirectional switching. (a) Free energy schematic visualizing the effect of applied electric field E at 170K with discommensuration (DC) density n as the order parameter. The dashed lines represent the overall energy landscape and oscillations correspond to the discrete annihilation and creation of triplet DC's (see bottom-left inset). $0 < E < E_c$ tilts the energy landscape towards C phase, allowing the system to gradually traverse intermediate metastable states with fewer DC's. $E > E_c$ abruptly shifts the system back into the NC state. (b) 1T-TaS$_2$ resistance vs. temperature and electric field for the sweeps shown in Fig. 3a showing the difference between forward and reverse transitions and metastable resistance states accessed. The schematics in the inset show growing C domains with increasing $R_{sd}$ in accordance with our model.

A conducting network of discommensurations (DC's) distinguishes the NC phase from the C phase[10–14], and so the areal density of DC's, n, can be used as an order parameter to describe the energetics of the NC-C transition[25], with n being zero (nonzero) in the C (NC) phase. The free energy landscape describing the initial state is shown by the black trace in the main panel of Fig. 4a. At 170K, the C phase is the thermodynamic ground state; however, the sample remains locked in the local NC minimum due to the activation barrier. Importantly, the energy is not a smooth function of n, as DC's must be nucleated and annihilated discretely in triplets in 1T-TaS$_2$ in order to conserve the phase around a DC core (see inset in Fig. 4a)[15]. As the DC's are more conducting than the C regions[26], the latter can be more highly polarized by an external electric field, E. Applying a finite E below a critical value, $E_c$, then lowers the energy more for states with smaller n, thus tilting the energy landscape to destabilize the NC state, as shown by the blue trace in Fig.

4a. For a given n, the energy further decreases linearly with E. This allows the system to gradually transition towards the C phase by traversing a series of local, metastable minima corresponding to a decreasing density of DCs.

This qualitatively captures the multi-step NC-to-C transitions observed in our experiment, which are summarized in Fig. 4b. Here, we have overlaid the temperature-dependent resistance $R_{sd}$ (in black) for the 5-8nm thick device and $R_{sd}$ vs. electric field E at 170K corresponding to the forward driving (in blue) from Fig. 3a for comparison. Applying relatively small fields in the NC phase allows access to multiple metastable states within the hysteresis region of the temperature-driven transition, which in light of our analysis and imaging, we now interpret as an incremental melting of the NC DC network that is distributed relatively uniformly across the sample (see insets in Fig. 4b).

When the system is near the C phase with higher resistance, application of $E > E_c \sim 0.7$V/µm causes an abrupt reverse transition back towards a lower resistance state. The green traces in Fig. 4b show two instances of this backward driving for comparison. In both cases, the 1T-TaS$_2$ resistance immediately after this transition is very close to that of the NC phase at higher temperature under equilibrium conditions, and so we have drawn the energy landscape in green in Fig. 4a to describe this state phenomenologically. The overall activation barrier between the starting and final states reflects the hysteresis between ramping field up and down, which is caused by a first-order phase transition. This may indicate additional nonlinear coupling between E and n that is beyond the scope of our model. Nonetheless, the original NC DC network is nucleated again in this free energy picture, resetting the system. The melting process is active again when ramping the field back to zero, however, which leaves the system in an intermediate state.

Abrupt voltage/current-induced transitions from the C phase into CDW states with higher conductivity have been previously observed under various conditions. In lateral devices with in-plane currents and fields, it has been attributed to a carrier-driven breakdown of the insulating state[3], as well as to the formation of conducting, textured domains[6], although Joule heating is also possible[4]. Indeed, mosaic-like metallic domains have been induced on bulk 1T-TaS$_2$ using the tip of a scanning tunneling microscope[27,28]. The vertical electric fields used in these studies are orders of magnitude larger than our $E_c$, however, which suggests a different mechanism. The NC resistance we observe above $E_c$ is also inconsistent with that of a random domain network. If the transition was instead induced by Joule heating above the warming transition temperature, we then expect through subsequent cooling as the voltage is ramped down that CDW domains would appear similar to that for the temperature-driven transition shown in Fig. 2. Our photocurrent images taken afterwards are also inconsistent with this scenario and thus point to other possibilities, such as a carrier-driven breakdown of the C phase.

Although further studies are needed to clarify the microscopic mechanism for the abrupt, first-order C-to-NC transition, our work has clearly demonstrated that the electrically induced transitions in both directions are spatially uniform on the micron scale, and paves the way for the use of 2D 1T-TaS$_2$ as a reversible multi-memristor material in future device applications.

**Acknowledgements:**

AWT acknowledges support from the US Army Research Office (W911NF-19-10267), Ontario Early Researcher Award (ER17-13-199), and the National Science and Engineering Research Council of Canada (RGPIN-2017-03815). This research was undertaken thanks in part to funding from the Canada First Research Excellence Fund. JO acknowledges support from Georg H. Endress Foundation. JJG, XL, WJL, and YPS appreciate the support of the National Key Research and Development Program under Contract No. 2016YFA0300404, the National Nature Science Foundation of China under Contracts No. 11674326, 11774351, and 11874357, and the Joint Funds of the National Natural Science Foundation of China and the Chinese Academy of Sciences' Large-Scale Scientific Facility under Contracts No. U1832141 and U1932217.